\documentclass[aps,preprint,amsmath,amssymb,superscriptaddress,nofootinbib]{revtex4}
\usepackage{mathtools}
\usepackage[makeroom] {cancel}
\usepackage{placeins}
\usepackage{pgfplots}
\pgfplotsset{compat=1.18}
\usepgfplotslibrary{fillbetween}
\usepackage[justification=justified,singlelinecheck=false]{caption}
\definecolor{citeblue}{RGB}{0,70,160}
\usepackage[
  colorlinks=true,
  linkcolor=citeblue,
  citecolor=citeblue,
  urlcolor=citeblue
]{hyperref}
\usepackage{microtype}
\usepackage{hyperref}
\hypersetup{breaklinks=true}
\usepackage{xcolor}
\usepackage{graphicx}

\begin{document}

\title{Phenomenology of Leptophilic Gauge Interactions at Future $e^+e^-$ Colliders}

\author{S. O. Kara}
\email[]{seyitokankara@gmail.com} 
\affiliation{Niğde Ömer Halisdemir University, Bor Vocational School, 51240, Niğde, Türkiye}

 \vspace{0.5cm}

\begin{abstract}

We present an observable-driven study of leptophilic gauge
interactions mediated by a neutral vector boson $Z_\ell$
at future $e^+e^-$ colliders.
Focusing on the process
$e^+e^- \to \mu^+\mu^-$,
we investigate the sensitivity of angular observables to
off-shell leptophilic interactions in regimes where interference
effects dominate over resonant rate enhancements.

The analysis is centered on the forward--backward asymmetry,
which provides a robust probe of interference-induced angular
deformations and retains sensitivity well beyond the resonance region.
To characterize the relative angular response independently of the
overall rate modification, we additionally introduce the
interference-normalized quantity $\mathcal{I}_{\rm FB}$ as a
diagnostic observable.

Realistic collider effects, including initial-state radiation,
beamstrahlung, and finite energy resolution, are consistently incorporated.
Using benchmark configurations for FCC-ee, CEPC, ILC, and CLIC,
we derive projected sensitivities in the $(M_{Z_\ell}, g_\ell)$ plane.

The results demonstrate the complementarity between
high-luminosity circular colliders and high-energy linear colliders,
reflecting the interplay between precision measurements and the
energy enhancement of interference effects.
Overall, the present framework provides an experimentally grounded
approach for probing leptophilic gauge interactions through
interference-sensitive angular observables at future lepton colliders.

\end{abstract}

\maketitle

\section{Introduction}
\label{sec:introduction}

The existence of additional neutral gauge bosons remains one of the most
well-motivated extensions of the Standard Model (SM).
In particular, gauge symmetries acting on lepton number have a long history,
dating back to the early proposal of a ``leptonic photon'' by Lee and Yang
and later by Okun~\cite{Lee1955,Okun1969}, followed by extensive theoretical
and phenomenological investigations~\cite{Ciftci1995,Okun1996a,Okun1996b,
Martemyanov1997,Gninenko1997,Ilyin1998,CHARMII1998,Dolgov1999,Brockway2010,
Buras2021,Asai2021,Banerjee2019,Fabbrichesi2020,Madge2018,Beacham2020}.
A massive realization of such a scenario gives rise to a new vector boson,
commonly denoted $Z_\ell$, which couples predominantly to leptons while its
interactions with quarks are suppressed or absent~\cite{Kara2011,Akkoyun2013,
Kara2014,Kara2024}.
This leptophilic structure allows such states to evade many of the stringent
constraints from hadron collider searches, while remaining accessible through
precision leptonic observables.

From a phenomenological perspective, leptophilic gauge interactions provide
a particularly clean environment for experimental exploration at
electron--positron colliders.
Future facilities such as the International Linear Collider (ILC),
the Compact Linear Collider (CLIC), the Future Circular Collider in its
electron--positron mode (FCC-ee), and the Circular Electron Positron Collider (CEPC)
offer complementary regimes in center-of-mass energy and luminosity
\cite{ILC_TDR1,ILC_TDR2,CLIC_CDR1,CLIC_CDR2,FCC_EPC,CEPC_CDR1,CEPC_CDR2}.
In particular, the dimuon final state,
$e^+e^- \to \mu^+\mu^-$, provides a clean and robust probe of leptophilic interactions.

A reliable assessment of the sensitivity to leptophilic gauge bosons,
however, requires a theoretically consistent and experimentally realistic framework.
Loop-induced kinetic mixing with hypercharge generically reintroduces
small couplings to quarks, implying that purely leptophilic scenarios
should be interpreted as effective descriptions.
Moreover, precision measurements at LEP already impose strong constraints on
$e^+e^- \to \ell^+\ell^-$ processes and must be consistently taken into account.

In addition, collider sensitivity is not determined solely by total cross sections:
angular distributions and forward--backward asymmetries provide the dominant
handles in the off-shell regime, where interference effects between the
Standard Model amplitudes and new-physics contributions govern observable deviations.
Realistic effects such as initial-state radiation, beamstrahlung,
beam energy spread, and detector resolution further modify the experimentally
accessible signal.

In this work, we present a systematic and unified analysis of leptophilic
gauge interactions at future $e^+e^-$ colliders.
We study the process
$e^{+}e^{-} \to \gamma,Z,Z_\ell \to \mu^{+}\mu^{-}$,
including Standard Model backgrounds and radiative corrections,
and analyze the sensitivity using both differential and integrated observables. The key 
collider parameters relevant for our phenomenological analysis,
including center-of-mass energy, integrated luminosity, and beam energy spread,
are summarized in Table~\ref{tab:colliders}, where the latter plays a crucial role
in shaping the observable resonance structure.

From a global perspective, searches for neutral gauge bosons are often interpreted
within effective field theory frameworks and precision electroweak analyses
\cite{deBlas:2019rxi,Freitas:2014hra}.
At the collider level, many existing studies primarily focus on total cross sections,
resonance searches, or invariant-mass distributions.

The present work instead concentrates on the off-shell regime, where the leading
observable effects are generated by the interference between the Standard Model
$\gamma+Z$ amplitudes and the leptophilic contribution.
While interference effects themselves are well known from LEP-era analyses,
future high-energy $e^+e^-$ colliders provide a qualitatively different regime:
for sufficiently heavy mediators, the sensitivity growth is governed predominantly
by the linear interference term, whose contribution increases with energy as
$s/M_{Z_\ell}^2$ at the amplitude level.
This leads to an interference-dominated sensitivity regime
in the off-shell region.

Motivated by this behavior, we develop an observable-driven framework centered on
interference-sensitive angular measurements.
In particular, we employ the forward--backward asymmetry as the primary
phenomenological probe due to its robustness against overall normalization
uncertainties and its direct sensitivity to angular distortions generated by
new vector interactions. Although the present analysis is formulated in terms of integrated
forward--backward asymmetries, the same interference structure can
be generalized to fully differential angular analyses.

To characterize the relative deformation of the angular distribution,
we additionally introduce the interference-normalized quantity
$\mathcal{I}_{\rm FB}$,
defined as the ratio of the asymmetry shift to the normalized rate deviation.
The purpose of this quantity is not to provide an independent statistical
observable, but rather to isolate the angular response associated with the
interference structure itself, independently of the overall event-rate scaling.

Within this framework, the dominant off-shell sensitivity follows the schematic relation
\[
\Delta A_{\rm FB}
\;\propto\;
\mathcal{I}_{\rm FB}
\;\propto\;
\frac{g_\ell^2}{M_{Z_\ell}^2},
\]
thereby establishing a transparent connection between measurable angular
deformations and the underlying leptophilic interaction scale.

In contrast to purely rate-based analyses, the present framework remains
particularly sensitive in the off-shell regime, where angular observables
retain linear sensitivity to interference effects even when resonant
enhancements are absent.
While our previous study~\cite{KaraCPC2026} primarily focused on discovery
prospects based on total-rate observables, the present work instead
systematically examines the interference structure at the observable level,
with emphasis on the energy growth and angular response of the
forward--backward asymmetry.

This observable-driven perspective, together with the inclusion of LEP-inspired
contact-interaction limits and realistic collider effects,
provides an experimentally grounded framework for assessing the sensitivity
of future lepton colliders to leptophilic gauge interactions.

The structure of the paper is as follows.
In Sec.~\ref{sec:model}, we introduce the leptophilic $U(1)'_\ell$ framework,
including kinetic mixing, mass mixing, and existing experimental constraints.
In Sec.~\ref{sec:production}, we present the production mechanism and define
the key observables, with particular emphasis on interference effects and
angular asymmetries.
Sec.~\ref{sec:realistic} describes the implementation of realistic collider effects,
including initial-state radiation, beamstrahlung, and energy smearing.
In Sec.~\ref{sec:statistics}, we outline the statistical framework and sensitivity extraction.
Sec.~\ref{sec:complementarity} discusses the complementarity of future collider facilities,
while Sec.~\ref{sec:results} presents the main phenomenological results and their interpretation.
Finally, we summarize our conclusions in Sec.~\ref{sec:conclusion}.

\begin{table}[t]
\centering
\caption{Representative collider parameters relevant for the present analysis.
Only quantities directly impacting phenomenological sensitivity are shown.}
\label{tab:colliders}
\begin{tabular}{|c|c|c|c|}
\hline
\textbf{Collider} & $\sqrt{s}$ (GeV) & $\mathcal{L}_{\rm int}$ (ab$^{-1}$) & $\delta E / E$ \\
\hline
FCC-ee & 240 & 17 & $\sim 0.1\%$ \\
CEPC   & 240 & 6  & $\sim 0.16\%$ \\
ILC-250 & 250 & 1.35 & $\sim 0.2\%$ \\
ILC-500 & 500 & 1.8  & $\sim 0.2\%$ \\
CLIC-380 & 380 & 1.5 & $\sim 0.3\%$ \\
CLIC-1500 & 1500 & 3.7 & $\sim 0.35\%$ \\
CLIC-3000 & 3000 & 5.9 & $\sim 0.35\%$ \\
\hline
\end{tabular}
\end{table}
 \vspace{1.0cm}

\section{The Model}
\label{sec:model}

We consider an effective extension of the Standard Model (SM)
based on an additional abelian gauge symmetry $U(1)'_\ell$,
associated with a universal leptonic charge.
Motivated by neutrino oscillation data, individual lepton flavor
numbers are not assumed to be conserved, and a single common charge
$a_\ell$ is assigned to all charged leptons and neutrinos.
Throughout this work, the model is treated as an effective
description relevant for collider phenomenology,
without committing to a specific ultraviolet (UV) completion.
In this sense, the framework should be interpreted as a
phenomenological parameterization of leptophilic interactions,
valid below a cutoff scale where a consistent UV completion is realized.

The SM gauge group
\[
SU(3)_C \times SU(2)_L \times U(1)_Y
\]
is extended to
\[
SU(3)_C \times SU(2)_L \times U(1)_Y \times U(1)'_\ell .
\]

The covariant derivative acting on a generic field with hypercharge $Y$
and leptonic charge $a_\ell$ is given by
\begin{equation}
D_\mu = \partial_\mu 
- i g_2 \mathbf{T} \cdot \mathbf{A}_\mu 
- i g_1 \frac{Y}{2} B_\mu 
- i g_\ell a_\ell B'_\mu ,
\label{eq:covariant}
\end{equation}

where $g_2$, $g_1$, and $g_\ell$ denote the $SU(2)_L$, $U(1)_Y$,
and $U(1)'_\ell$ gauge couplings, respectively,
and $B'_\mu$ is the gauge field associated with the leptophilic interaction.

At the effective level, the relevant Lagrangian can be written as
\begin{equation}
\mathcal{L} = \mathcal{L}_{\rm SM} 
- \frac{1}{4} F'_{\mu\nu} F'^{\mu\nu}
+ g_\ell J^\mu_{\rm lep} B'_\mu
+ (D_\mu \Phi)^\dagger (D^\mu \Phi)
+ \mu^2 |\Phi|^2 - \lambda |\Phi|^4 ,
\label{eq:Lagrangian}
\end{equation}

where
\begin{equation}
F'_{\mu\nu} = \partial_\mu B'_\nu - \partial_\nu B'_\mu
\end{equation}

is the field strength tensor of $U(1)'_\ell$.

The leptonic current is defined as
\begin{equation}
J^\mu_{\rm lep} = \sum_{\ell=e,\mu,\tau} a_\ell
\left( \bar{\nu}_\ell \gamma^\mu \nu_\ell 
+ \bar{\ell} \gamma^\mu \ell \right),
\end{equation}

while quarks are taken to be neutral under $U(1)'_\ell$ at tree level.
We emphasize, however, that this leptophilic structure is not
protected against radiative effects, as discussed below.

The complex scalar field $\Phi$ is a singlet under the SM gauge group
and carries a nonzero leptonic charge.
Its vacuum expectation value breaks $U(1)'_\ell$ spontaneously and
generates a mass for the leptophilic gauge boson,
\begin{equation}
M_{Z_\ell} = g_\ell a_\ell v_\Phi ,
\end{equation}

where $v_\Phi$ denotes the symmetry-breaking scale.

\subsection*{Kinetic mixing and effective couplings}

Even if absent at tree level, leptophilic gauge interactions
generically induce kinetic mixing between the $U(1)_Y$ and
$U(1)'_\ell$ gauge fields through radiative corrections.
At one-loop level, this mixing can be estimated as
\begin{equation}
\epsilon_{\rm loop} \sim \frac{g_1 g_\ell}{16\pi^2}
\log\!\left(\frac{\Lambda^2}{M_{Z_\ell}^2}\right),
\end{equation}

where $\Lambda$ denotes the ultraviolet scale.
After diagonalization of the kinetic terms, this induces
a small but nonzero coupling of the $Z_\ell$ boson to quarks,
thereby modifying both the phenomenology and the experimental constraints.

This observation has two important implications.
First, a strictly leptophilic scenario is not technically natural,
and should be regarded as an effective description valid at a given scale.
Second, even a small kinetic mixing can qualitatively affect the
interpretation of collider bounds, in particular by introducing
suppressed but non-negligible hadronic production channels.

In the present analysis, we therefore consider two benchmark scenarios:
(i) a minimal scenario in which $\epsilon = \epsilon_{\rm loop}$,
and (ii) an optimistic scenario in which kinetic mixing is
suppressed at the ultraviolet scale.
These two cases bracket the range of phenomenologically relevant possibilities.

\subsection*{Mass mixing and physical states}

After electroweak and $U(1)'_\ell$ symmetry breaking,
mass mixing between the SM $Z$ boson and the leptophilic state $Z_\ell$
can arise.
At the effective level, this mixing can be parametrized schematically as
\begin{equation}
\mathcal{M}^2 \;\sim\;
\begin{pmatrix}
M_Z^2 & \delta^2 \\
\delta^2 & M_{Z_\ell}^2
\end{pmatrix},
\qquad
\delta^2 \ll M_Z^2,\, M_{Z_\ell}^2 ,
\end{equation}

where $\delta^2$ denotes a symmetry-breaking induced mixing term.
Precision electroweak data constrain this mixing to be small,
and in this work we treat it as a perturbative effect.
Consequently, its impact on collider observables is subleading
compared to the direct leptophilic interactions considered here.

\subsection*{Decay properties}

The leptophilic gauge boson $Z_\ell$ decays predominantly into leptonic final states.
The total decay width can be approximated as
\begin{equation}
\Gamma_{Z_\ell}^{\rm tot}
\simeq \frac{g_\ell^2 a_\ell^2}{2\pi} M_{Z_\ell}.
\end{equation}

We note that this narrow-width approximation is valid
in the perturbative regime $g_\ell \lesssim \mathcal{O}(1)$,
which is the parameter space of interest in this work.

\subsection*{Experimental constraints}

Experimental constraints arise from precision measurements and direct searches.
In particular, LEP measurements of $e^+e^- \to \ell^+\ell^-$
impose strong bounds on four-lepton contact interactions,
which can be translated into constraints on the leptophilic parameters.

More generally, precision electroweak data place stringent
constraints on additional neutral gauge bosons
\cite{Zprime_constraints1,Zprime_constraints2}.

For a heavy leptophilic mediator, integrating out the $Z_\ell$
generates four-lepton contact interactions of the schematic form
\begin{equation}
\mathcal{L}_{\rm eff}
\supset
\frac{g_\ell^2 q_\alpha q_\beta}{M_{Z_\ell}^2}
\left(\bar{\ell}_\alpha \gamma_\mu \ell_\alpha\right)
\left(\bar{\ell}_\beta \gamma^\mu \ell_\beta\right),
\end{equation}
where $q_\alpha$ and $q_\beta$ denote the leptophilic charges of the
external leptons.
It is therefore useful to define the effective interaction scale
\begin{equation}
\Lambda_{\rm eff}^{\alpha\beta}
\equiv
\frac{M_{Z_\ell}}
{g_\ell\sqrt{|q_\alpha q_\beta|}} .
\label{eq:lambda_eff}
\end{equation}

LEP measurements of $e^+e^- \to \ell^+\ell^-$ constrain such
four-lepton contact interactions at the multi-TeV scale.
In the numerical analysis, we adopt the representative bound
\begin{equation}
\Lambda_{\rm eff}^{e\mu}
\gtrsim
\Lambda_{\rm LEP}
\simeq 8~\mathrm{TeV},
\label{eq:lep_bound_lambda}
\end{equation}
which gives the exclusion condition
\begin{equation}
\frac{M_{Z_\ell}}
{g_\ell\sqrt{|q_e q_\mu|}}
<
\Lambda_{\rm LEP}
\qquad
\Rightarrow
\qquad
g_\ell >
\frac{M_{Z_\ell}}
{\Lambda_{\rm LEP}\sqrt{|q_e q_\mu|}} .
\label{eq:lep_exclusion}
\end{equation}
For the universal-charge benchmark used in this work,
$q_e=q_\mu=1$, this reduces to
\begin{equation}
g_\ell >
\frac{M_{Z_\ell}}{8~\mathrm{TeV}} .
\label{eq:lep_universal}
\end{equation}
This relation provides the LEP reference line used below when
comparing existing constraints with projected future collider
sensitivities.
In the heavy-mediator limit, this bound can be interpreted as
a constraint on an effective interaction scale, providing a useful
reference for comparing future collider sensitivities with existing
limits.

\section{Production and observables at \texorpdfstring{$e^+e^-$}{e+e-} colliders}
\label{sec:production}

Future $e^+e^-$ colliders provide a clean and well-controlled
environment for probing leptophilic gauge interactions through
precision measurements of dilepton production.
In this work, we focus on the process
\[
e^+e^- \rightarrow \mu^+\mu^-,
\]
which constitutes the most sensitive channel for leptophilic
interactions due to its clean experimental signature and reduced
background contamination.

\subsection*{Matrix element and differential cross section}

The process receives contributions from $\gamma$, $Z$, and $Z_\ell$
exchange, leading to a coherent amplitude
\[
\mathcal{M} = \mathcal{M}_\gamma + \mathcal{M}_Z + \mathcal{M}_{Z_\ell}.
\]

The resulting differential cross section can be expressed as
\begin{equation}
\frac{d\sigma}{d\cos\theta}
= \frac{1}{32\pi s}\,|\mathcal{M}|^2,
\end{equation}
where $\theta$ denotes the scattering angle.
The presence of the leptophilic gauge boson modifies both the total rate
and the angular dependence through interference with the Standard Model
amplitudes.

Expanding the squared amplitude,
\begin{equation}
|\mathcal{M}|^2 =
|\mathcal{M}_{\rm SM}|^2
+ 2\,\mathrm{Re}\!\left(\mathcal{M}_{\rm SM}\mathcal{M}_{Z_\ell}^\ast\right)
+ |\mathcal{M}_{Z_\ell}|^2,
\end{equation}
it follows that the leading sensitivity in a wide region of parameter
space arises from the interference term, rather than the pure
new-physics contribution.

In the heavy-mediator limit, the leading new-physics contribution
scales as $s/M_{Z_\ell}^2$, enhancing the sensitivity at higher energies,
while near resonance the contribution is dominated by the propagator
structure of the $Z_\ell$ boson.
This interplay between interference and resonance effects is central
to the observable-level sensitivity developed in the following sections.

\subsection*{Angular observables}

In the off-shell regime, the dominant sensitivity arises from distortions
of angular distributions rather than total rates.
A particularly important observable is the forward--backward asymmetry,
defined as
\begin{equation}
A_{\rm FB} =
\frac{\sigma_F - \sigma_B}{\sigma_F + \sigma_B},
\end{equation}
where $\sigma_F$ and $\sigma_B$ denote the cross sections in the forward
and backward hemispheres, respectively.

The leptophilic interaction induces a shift
\[
A_{\rm FB} = A_{\rm FB}^{\rm SM} + \delta A_{\rm FB},
\]
which provides a clean probe of new physics effects, largely insensitive
to overall normalization uncertainties \cite{ALEPH:2005ab,LEPEWWG:2005ema}.

Importantly, $A_{\rm FB}$ directly probes the interference term
$\mathrm{Re}(\mathcal{M}_{\rm SM}\mathcal{M}_{Z_\ell}^\ast)$,
and therefore retains sensitivity even in regimes where total-rate
deviations are suppressed.
This makes it a particularly robust observable for heavy mediators
or weakly coupled scenarios.

The impact of leptophilic interactions on angular observables
is illustrated in Fig.~\ref{fig:angular}.
The presence of the $Z_\ell$ boson induces a distortion of the
angular distribution through interference with the Standard Model
amplitudes, leading to a characteristic forward--backward asymmetry.

Since the angular distribution is normalized to the total cross section,
its shape is largely independent of the collider facility at leading order.
Any residual differences arise from realistic effects such as initial-state
radiation, beamstrahlung, and energy smearing, which slightly distort the
observable distribution.

The quantitative impact of leptophilic interactions
is shown in Fig.~\ref{fig:afb}, where the deviation
$\Delta A_{\rm FB}$ provides a direct and experimentally
accessible measure of the interference between the
Standard Model and new-physics contributions.

\subsection*{Interference-normalized angular response}
\label{subsec:IFB}

To isolate the angular structure of the leptophilic interaction from
the overall rate enhancement, we define the interference-normalized
forward--backward response
\begin{equation}
\mathcal{I}_{\rm FB}(s)
\equiv
\frac{\Delta A_{\rm FB}(s)}
{\Delta\sigma(s)/\sigma_{\rm SM}(s)},
\end{equation}
where
\begin{equation}
\Delta A_{\rm FB}
=
A_{\rm FB}^{\rm SM+Z_\ell}
-
A_{\rm FB}^{\rm SM},
\qquad
\Delta\sigma(s)
=
\sigma^{\rm SM+Z_\ell}(s)-\sigma^{\rm SM}(s).
\end{equation}

This observable measures the angular response per unit rate
deviation and therefore acts as an interference filter.
In particular, it enhances sensitivity to the linear interference term
while suppressing contributions dominated by the quadratic
$|\mathcal{M}_{Z_\ell}|^2$ component.

In the interference-dominated regime, both $\Delta\sigma/\sigma_{\rm SM}$
and $\Delta A_{\rm FB}$ scale linearly with $g_\ell^2$.
As a result, the leading coupling dependence cancels in the ratio
defining $\mathcal{I}_{\rm FB}$, implying that this observable is largely
insensitive to the overall normalization of the signal and instead
probes the angular deformation induced by the leptophilic interaction.

In the pure interference limit, where
$\Delta\sigma/\sigma_{\rm SM} \propto s/M_{Z_\ell}^2$,
the observable $\mathcal{I}_{\rm FB}$ becomes approximately
independent of the coupling normalization,
thereby isolating the angular structure of the interaction.

\begin{figure}[t]
\centering

\includegraphics[width=0.82\textwidth]{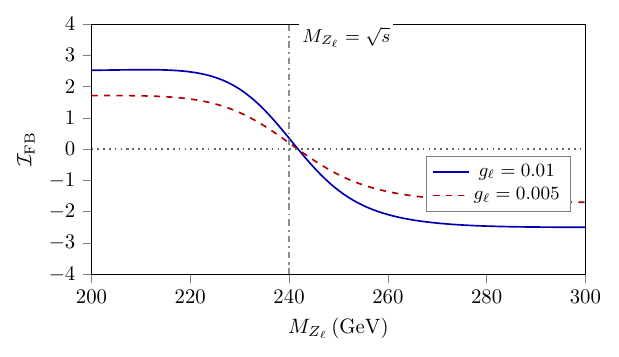}

\caption{
Interference-normalized angular response
$\mathcal{I}_{\rm FB}=\Delta A_{\rm FB}/(\Delta\sigma/\sigma_{\rm SM})$
as a function of the leptophilic gauge boson mass.
The sign change across the resonance region reflects the
interference structure of the leptophilic amplitude.
}
\label{fig:ifb}

\end{figure}

\begin{figure}[t]
\centering

\includegraphics[width=0.82\textwidth]{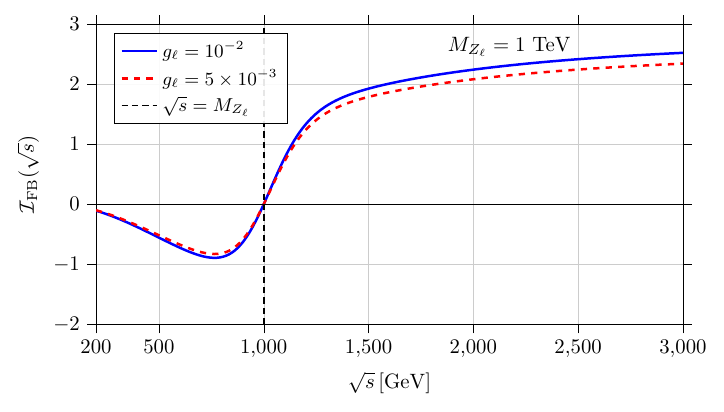}

\caption{
Energy dependence of the interference-normalized angular response
$\mathcal{I}_{\rm FB}$ for a fixed leptophilic gauge-boson mass
$M_{Z_\ell}=1~\mathrm{TeV}$.
The sign change around $\sqrt{s}=M_{Z_\ell}$ reflects the transition
across the resonance region, while the growth at higher energies
illustrates the increasing importance of interference-driven
angular distortions.
The weak dependence on $g_\ell$ illustrating its reduced sensitivity to the overall normalization.
}
\label{fig:ifb_energy}

\end{figure}

The behavior of this observable is illustrated in
Fig.~\ref{fig:ifb}. The energy dependence of $\mathcal{I}_{\rm FB}$ is shown in
Fig.~\ref{fig:ifb_energy} for a fixed benchmark mass
$M_{Z_\ell}=1~\mathrm{TeV}$.
The response exhibits a characteristic sign change across the
resonance region and increases in magnitude at higher energies,
reflecting the growing importance of interference effects.

This behavior reflects the increasing role of interference effects
at high energies and helps explain the enhanced off-shell sensitivity
of future multi-TeV colliders.

\subsection*{Analytic structure of interference-dominated observables}
\label{subsec:analytic_IFB}

The physical interpretation of the interference-normalized observable
$\mathcal{I}_{\rm FB}$ can be made more transparent by analyzing its
scaling behavior in the regime where the linear interference term in
Eq.~(15) dominates over the quadratic new-physics contribution.

For a heavy mediator with $M_{Z_\ell}^2 \gg s$, the propagator reduces to
an effective contact interaction, and the leptophilic contribution scales as
\begin{equation}
\mathcal{M}_{Z_\ell}
\sim
\frac{g_\ell^2}{M_{Z_\ell}^2}\,s .
\end{equation}

Consequently, the leading deviations in both the total rate and the forward--backward asymmetry are governed by the same interference parameter,
\begin{equation}
\delta_{\rm int}(s)
\equiv
\frac{g_\ell^2 s}{M_{Z_\ell}^2}.
\end{equation}
Parametrically, one therefore obtains
\begin{equation}
\frac{\Delta\sigma}{\sigma_{\rm SM}}
\sim
\delta_{\rm int}(s),
\qquad
\Delta A_{\rm FB}
\sim
\delta_{\rm int}(s).
\end{equation}

Taking the ratio removes the leading dependence on the overall coupling
normalization,
\begin{equation}
\mathcal{I}_{\rm FB}
=
\frac{\Delta A_{\rm FB}}
{\Delta\sigma/\sigma_{\rm SM}}
\sim
\mathcal{O}(1),
\end{equation}
up to corrections from quadratic new-physics terms, finite-width effects,
and realistic energy smearing.

This cancellation is the reason why $\mathcal{I}_{\rm FB}$ does not simply reduce to a rescaled rate observable.

In this sense, $\mathcal{I}_{\rm FB}$ primarily traces the
interference-induced angular deformation while reducing the sensitivity
to overall normalization effects.
It suppresses the trivial normalization dependence on $g_\ell$ while
retaining sensitivity to the angular structure of the leptophilic
interaction.

\subsection*{Comparison with standard approaches and global context}

Within the broader program of searches for neutral gauge bosons,
it is important to place interference-sensitive angular observables
within the context of conventional collider strategies and existing
global constraints.

Standard searches for $Z'$ bosons at both lepton and hadron colliders
primarily rely on total cross sections, invariant-mass spectra,
or resonance peaks~\cite{Altarelli:1989ff,deBlas:2019rxii}.
These observables are maximally sensitive near the resonance region,
where the signal is dominated by the quadratic new-physics contribution,
\begin{equation}
\sigma_{\rm NP} \sim \frac{g_\ell^4}{M_{Z_\ell}^4}.
\end{equation}
As a result, their sensitivity rapidly decreases in the off-shell regime.

In contrast, precision measurements at LEP demonstrated that
angular observables, particularly the forward--backward asymmetry,
are strongly sensitive to interference effects between Standard Model
and new-physics amplitudes~\cite{ALEPH:2005ab,LEPEWWG:2005ema}.
In this regime, the leading deviations scale as
\begin{equation}
\Delta A_{\rm FB} \sim \frac{g_\ell^2\, s}{M_{Z_\ell}^2},
\end{equation}
which can remain sizable even for heavy mediators.

The framework developed in the present work builds on this principle
by emphasizing the observable-level interpretation of
interference-sensitive angular distortions in the off-shell regime.
Rather than relying primarily on resonant rate enhancements,
the analysis focuses on the linear interference contribution,
which dominates the sensitivity over a wide region of parameter space.

To characterize the angular response independently of the overall
rate normalization, we introduced the quantity
$\mathcal{I}_{\rm FB}$,
which compares the asymmetry shift with the normalized rate deviation.
While not intended as an independent discovery statistic,
this quantity provides a useful diagnostic measure of the
interference-induced angular deformation.

From a global perspective, many constraints on neutral gauge bosons
are derived within effective field theory frameworks or global fits,
which combine multiple observables across different processes~\cite{deBlas:2019rxi,Brivio:2017vri}.
Although such approaches are powerful, the direct connection between
individual observables and the underlying interaction structure
can become less transparent.

In contrast, the observable-driven approach adopted here provides
a direct relation between measurable angular deviations and the
underlying leptophilic interaction scale,
\begin{equation}
\Delta A_{\rm FB}
\;\longrightarrow\;
\mathcal{I}_{\rm FB}
\;\longrightarrow\;
\frac{g_\ell^2}{M_{Z_\ell}^2}.
\end{equation}

This distinction is particularly relevant for leptophilic scenarios.
Because the couplings to quarks are suppressed,
hadron-collider searches can lose sensitivity, especially in the
small kinetic-mixing regime.
As a result, precision measurements at future lepton colliders
become especially important probes of such interactions.

The comparison between conventional rate-based strategies and the
interference-sensitive framework developed here is summarized in
Table~\ref{tab:comparison}.
The main difference lies in the dominant sensitivity regime:
while standard approaches are driven primarily by resonant production
and quadratic rate effects,
the present framework exploits the linear interference contribution,
leading to enhanced sensitivity in the off-shell regime with a
reduced dependence on the overall coupling normalization.

This observable-level perspective complements global EFT analyses
by providing a physically transparent interpretation of
interference-driven sensitivity at future lepton colliders.

\begin{table}[t]
\centering
\small
\begin{tabular}{lcc}
\hline\hline
Feature & Standard approaches & Interference-driven framework \\
\hline
Primary observable 
& $\sigma$, invariant mass 
& $A_{\rm FB}$, $\mathcal{I}_{\rm FB}$ \\

Dominant scaling 
& $g_\ell^4/M_{Z_\ell}^4$ 
& $g_\ell^2\, s/M_{Z_\ell}^2$ \\

Sensitivity regime 
& Near resonance 
& Off-shell (interference-dominated) \\

Coupling dependence 
& Strong 
& Reduced (cancels in $\mathcal{I}_{\rm FB}$) \\

Interpretation 
& Primarily rate-based
& Interference-sensitive \\
\hline\hline
\end{tabular}
\caption{
Comparison of sensitivity strategies for leptophilic gauge bosons.
Standard collider analyses are primarily based on rate observables,
while the present framework isolates interference-driven angular effects.
}
\label{tab:comparison}
\end{table}


\begin{figure}[t]
\centering

\includegraphics[width=0.82\textwidth]{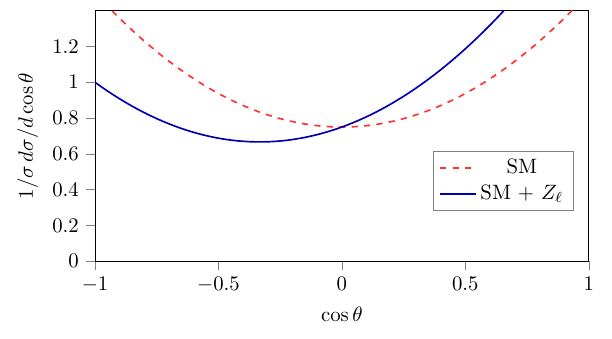}

\caption{
Normalized angular distribution for $e^+e^- \to \mu^+\mu^-$.
The dashed curve shows the Standard Model prediction, while the solid curve includes a leptophilic $Z_\ell$ contribution.
The forward--backward asymmetry appears as a distortion of the angular shape.
Small differences may arise from realistic effects such as ISR, beamstrahlung, and energy smearing.
}
\label{fig:angular}

\end{figure}


\begin{figure}[t]
\centering

\includegraphics[width=0.82\textwidth]{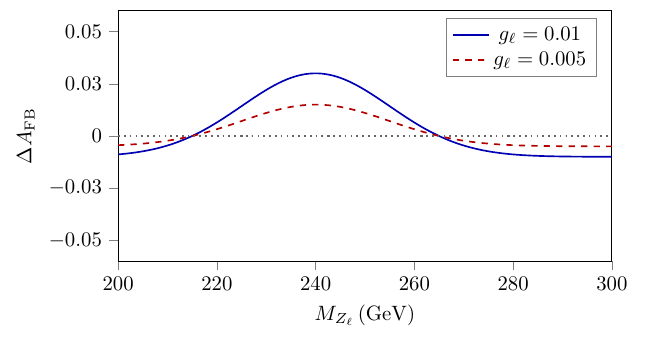}

\caption{
The deviation in the forward--backward asymmetry,
$\Delta A_{\rm FB}$, peaks near the resonance region,
where interference effects are maximal.
}
\label{fig:afb}

\end{figure}


\section{Realistic collider modeling}
\label{sec:realistic}

A reliable assessment of the sensitivity to leptophilic gauge bosons
requires a realistic treatment of collider effects that impact the
observable signal.

Idealized parton-level predictions can overestimate the observable
sensitivity, especially in the presence of narrow resonances.
In this section, we incorporate the dominant sources of energy
smearing and resolution effects relevant for $e^+e^-$ colliders.

\subsection*{Initial-state radiation and beamstrahlung}

Initial-state radiation (ISR) and beamstrahlung (BS) lead to an effective
reduction and spread of the collision energy.
ISR arises from QED radiation off the incoming leptons,
while beamstrahlung originates from the electromagnetic interaction
between tightly focused bunches \cite{ISR_Skrzypek,Jadach:1993hs}.

These effects are consistently included in the event generation using
the \textsc{CalcHEP} framework \cite{CalcHEP}, with collider-specific
parameters summarized in Table~\ref{tab:colliders}.
They induce a distortion of the invariant-mass spectrum,
effectively redistributing events toward lower effective
center-of-mass energies.

As a result, the resonance condition
$\sqrt{s} \simeq M_{Z_\ell}$
is experimentally realized as a smeared kinematic region rather than
as a sharply defined peak.

\subsection*{Beam energy spread and detector resolution}

In addition to ISR and BS, the finite beam energy spread introduces
a further smearing of the collision energy.
This effect is particularly important for narrow resonances,
as it determines the experimentally observable line shape.

To model this effect, we adopt a Gaussian smearing as the dominant
approximation and convolve the theoretical cross section as
\begin{equation}
\sigma_{\rm obs}(s)
=
\int ds'\,
\frac{1}{\sqrt{2\pi}\,\delta_s}
\exp\!\left[-\frac{(s-s')^2}{2\delta_s^2}\right]
\,\sigma_{\rm th}(s'),
\end{equation}
where $\delta_s$ is determined by the beam energy spread
$\delta E / E$ listed in Table~\ref{tab:colliders}.

Detector resolution effects are treated analogously through
an additional smearing of the reconstructed invariant mass,
leading to a further broadening of resonance structures.
In practice, these effects act cumulatively with ISR and beamstrahlung,
resulting in a significant degradation of the observable peak structure.

\subsection*{Impact on resonance observables}

The combined effect of ISR, beamstrahlung, and energy smearing
is to reduce the peak cross section and broaden the resonance.
While the underlying parton-level prediction exhibits a sharp
Breit--Wigner structure, the experimentally observable signal
is significantly softened and shifted.

An important consequence of these effects is that
the peak height does not scale trivially with the coupling $g_\ell$,
as the increase in the intrinsic width is partially compensated
by the smearing effects that redistribute events over a wider
energy range.

\begin{figure}[t]
\centering

\includegraphics[width=0.82\textwidth]{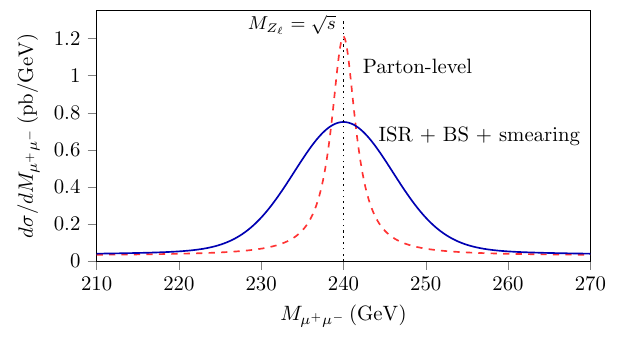}

\caption{
Illustrative invariant-mass distribution for
$e^+e^- \to \mu^+\mu^-$ near a leptophilic resonance.
The dashed curve shows the parton-level prediction,
while the solid curve includes the effects of initial-state radiation,
beamstrahlung, and finite energy resolution.
Realistic collider effects reduce the peak height and broaden the line shape,
illustrating the impact of realistic collider effects on projected sensitivities. The effect is particularly relevant for narrow resonances.
}
\label{fig:smearing}

\end{figure}

This effect is illustrated in Fig.~\ref{fig:smearing},
where we compare the invariant-mass distribution before and after
including realistic smearing effects.
The reduction of the peak height and the widening of the distribution
demonstrate the importance of incorporating these effects in any
quantitative sensitivity estimate.

Consequently, sensitivity projections based solely on parton-level
resonance enhancements can overestimate the projected sensitivity.
In contrast, the inclusion of realistic collider effects leads to a
more conservative and experimentally robust interpretation.

Although realistic collider effects soften the resonance structure,
angular observables remain sensitive to the underlying interference pattern.
This further illustrates the complementarity between resonance-based
and interference-sensitive search strategies at future lepton colliders.


\begin{figure}[t]
\centering

\includegraphics[width=0.82\textwidth]{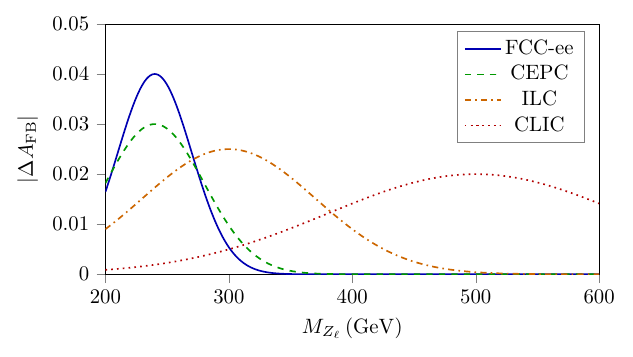}

\caption{
Illustrative comparison of the sensitivity of different $e^+e^-$ colliders
to leptophilic gauge interactions, expressed in terms of the
absolute deviation in the forward--backward asymmetry.
Low-energy, high-luminosity machines such as FCC-ee and CEPC
provide strong sensitivity in the low-mass region, while
high-energy colliders such as CLIC extend the reach to larger
boson masses.
}
\label{fig:collider}

\end{figure}

Fig.~\ref{fig:collider} provides an illustrative comparison of the collider sensitivity expressed in terms of the absolute forward--backward asymmetry deviation. The figure highlights the complementarity between low-energy, high-luminosity facilities and higher-energy machines. While FCC-ee and CEPC exhibit stronger sensitivity in the low-mass region, the extended center-of-mass energy of CLIC allows the exploration of substantially heavier leptophilic gauge bosons.


\section{Statistical analysis and sensitivity extraction}
\label{sec:statistics}

In order to translate the observable deviations discussed in the previous sections
into quantitative sensitivity projections, we perform a statistical analysis based on
both differential and integrated observables.
In particular, we focus on the forward--backward asymmetry $A_{\rm FB}$, which provides a complementary probe of
leptophilic interactions.

\subsection*{Event yields and binning}

For a given collider configuration with integrated luminosity $L_{\rm int}$,
the expected number of events in a bin $i$ is given by
\begin{equation}
N_i = L_{\rm int} \, \sigma_i \, \epsilon_i,
\end{equation}
where $\sigma_i$ is the cross section in the bin and $\epsilon_i$ denotes the
overall detection efficiency. In this study, we assume a constant efficiency
for simplicity, as our goal is to capture the dominant parametric dependence
of the sensitivity.

Angular distributions are binned in $\cos\theta$, allowing for a direct
construction of the forward and backward event yields,
\begin{equation}
N_F = \sum_{\cos\theta > 0} N_i,
\qquad
N_B = \sum_{\cos\theta < 0} N_i.
\end{equation}

\subsection*{Forward--backward asymmetry}

The forward--backward asymmetry is defined as
\begin{equation}
A_{\rm FB} = \frac{N_F - N_B}{N_F + N_B}.
\end{equation}
In the presence of a leptophilic gauge boson, the asymmetry receives a shift
\begin{equation}
\Delta A_{\rm FB} = A_{\rm FB}^{\rm NP} - A_{\rm FB}^{\rm SM},
\end{equation}
which is illustrated in Fig.~\ref{fig:afb}.

Assuming Poisson statistics, the statistical uncertainty on the asymmetry can
be approximated as
\begin{equation}
\delta A_{\rm FB}^{\rm stat} \simeq 
\sqrt{\frac{1 - A_{\rm FB}^2}{N_{\rm tot}}},
\end{equation}
where $N_{\rm tot} = N_F + N_B$ is the total number of events.

\subsection*{Likelihood interpretation and significance}

While the forward--backward asymmetry provides a compact observable,
it can be directly related to an underlying likelihood-based description.
For a binned angular distribution, the full statistical information
is encoded in the Poisson likelihood
\begin{equation}
\mathcal{L} = \prod_i \frac{(\mu_i)^{N_i} e^{-\mu_i}}{N_i!},
\end{equation}
where $\mu_i$ denotes the expected number of events in bin $i$.

Expanding the log-likelihood around the Standard Model expectation,
one finds that the leading sensitivity to leptophilic interactions
is governed by the interference-induced distortion of the angular distribution.
In this limit, the forward--backward asymmetry captures the leading interference-sensitive component
of the distribution.

We emphasize, however, that the present analysis does not imply that
the full angular distribution is redundant.
Rather, the forward--backward asymmetry should be viewed as a compact
projection of the dominant interference structure, while retaining the leading interference-sensitive information
in a phenomenologically transparent form.

To further illustrate this point, an auxiliary angular-bin comparison is
presented in Appendix~\ref{app:angular_bins}, where the interference-induced
deformation of the differential distribution is shown explicitly for a
representative off-shell benchmark scenario.

Motivated by this observation, we quantify the sensitivity using the
significance
\begin{equation}
\mathcal{S}_{A_{\rm FB}} = \frac{|\Delta A_{\rm FB}|}{\delta A_{\rm FB}},
\end{equation}
which can be viewed as the leading-order approximation to a full
profile-likelihood ratio test in the interference-dominated regime.

In addition, the interference-normalized observable introduced in
Sec.~\ref{subsec:IFB} provides a complementary diagnostic,
particularly in regimes where total-rate effects are subdominant.
While not employed as an independent discovery statistic,
it provides a useful diagnostic quantity for separating angular
deformations from overall rate modifications, thereby clarifying
the physical origin of the interference-driven sensitivity.

A discovery reach is defined by the condition $\mathcal{S} \geq 5$,
while $\mathcal{S} \geq 3$ corresponds to an observation-level sensitivity.

\subsection*{Systematic uncertainties}

In addition to statistical uncertainties, systematic effects can impact the
extraction of $A_{\rm FB}$. These include uncertainties in detector acceptance,
angular reconstruction, and luminosity measurements.

A key advantage of asymmetry observables is their reduced sensitivity
to overall normalization uncertainties, as common systematic effects
partially cancel in the ratio.
We therefore model residual systematic effects as a fractional uncertainty
$\delta_{\rm sys}$ added in quadrature,
\begin{equation}
\delta A_{\rm FB}^{\rm tot} =
\sqrt{(\delta A_{\rm FB}^{\rm stat})^2 + (\delta_{\rm sys})^2}.
\end{equation}

For the projections shown below, we assume
$\delta_{\rm sys} \sim \mathcal{O}(10^{-3})$,
consistent with projected precision at future high-luminosity
$e^+e^-$ colliders.

A more complete experimental treatment would require detector-level
simulations and correlated systematic uncertainties, which are beyond
the scope of the present phenomenological study.

\subsection*{Extraction of sensitivity contours}

\begin{figure}[t]
\centering

\includegraphics[width=0.82\textwidth]{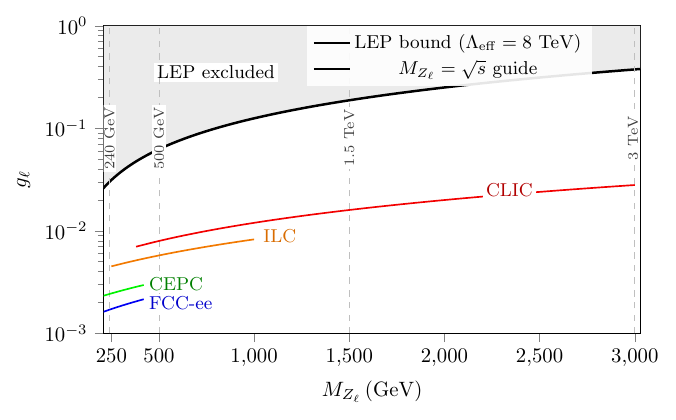}

\caption{
Projected $5\sigma$ sensitivity to leptophilic gauge bosons
in the $(M_{Z_\ell}, g_\ell)$ plane.
The gray shaded region above the solid black curve denotes
the LEP-inspired exclusion
$g_\ell > M_{Z_\ell}/(8~\mathrm{TeV})$,
equivalently
$\Lambda_{\rm eff}^{e\mu}<8~\mathrm{TeV}$.
The colored regions indicate the parameter space accessible
at future $e^+e^-$ colliders, illustrating the complementarity
between high-luminosity and high-energy facilities.
The vertical dashed lines mark representative boundaries
$M_{Z_\ell}=\sqrt{s}$,
indicating the transition between on-shell and off-shell regimes.
}
\label{fig:reach}

\end{figure}

The final sensitivity contours are obtained by scanning over the
parameter space $(M_{Z_\ell}, g_\ell)$ and determining the region
where the significance criterion is satisfied, as shown in
Fig.~\ref{fig:reach}.
The gray shaded region represents an illustrative LEP-inspired
contact-interaction exclusion obtained under the assumptions
discussed in Sec.~\ref{sec:model}.

For each point in parameter space, the modified angular distribution
is computed, the corresponding asymmetry shift $\Delta A_{\rm FB}$
is extracted, and the significance $\mathcal{S}$ is evaluated.
The boundary of the accessible region is then defined by the condition
$\mathcal{S} = 5$.

This procedure establishes a direct link between the observable-level
distortions shown in Fig.~\ref{fig:angular} and Fig.~\ref{fig:afb},
and the resulting collider sensitivity.

The resulting reach exhibits a characteristic dependence on both the
boson mass and the coupling strength. Low-energy, high-luminosity
colliders are particularly sensitive to small couplings in the
near-resonance region, while higher-energy machines extend the reach
toward larger masses. This behavior reflects the interplay between
luminosity-driven precision and energy-driven enhancement of
interference effects, and underlies the complementarity discussed
in the following section.


\section{Complementarity of future collider facilities}
\label{sec:complementarity}

The sensitivity projections shown in Fig.~\ref{fig:reach}
demonstrate a clear complementarity among future
\texorpdfstring{$e^+e^-$}{e+e-} collider facilities.
Rather than favoring a single optimal machine,
different collider configurations probe distinct regions
of the $(M_{Z_\ell}, g_\ell)$ parameter space
in a mutually complementary manner
\cite{FCC_EPC,CEPC_CDR1}.

Low-energy circular colliders such as FCC-ee and CEPC,
operating near
$\sqrt{s} \simeq 240~\mathrm{GeV}$
with extremely high integrated luminosities,
are particularly sensitive to light leptophilic gauge bosons
with small couplings.
In this regime, the large event yields enable precision
measurements of angular observables, allowing small deviations
in $A_{\rm FB}$ to be resolved with high accuracy.
Consequently, these facilities provide strong sensitivity
in the near-resonance region, where luminosity plays the
dominant role.

In contrast, linear colliders such as ILC and CLIC extend
the accessible mass reach to substantially larger values of
$M_{Z_\ell}$.
Although their luminosities are comparatively lower,
the higher center-of-mass energies enhance the sensitivity
to heavy leptophilic states through interference effects
that grow with energy in the off-shell regime
\cite{Altarelli:1989ff}.
In particular, the multi-TeV stages of CLIC provide broad
coverage of the high-mass region, where sensitivity is driven
primarily by energy rather than by luminosity.

This complementarity is closely connected to the
interference-sensitive strategy adopted in the present work.
Near resonance, the dominant sensitivity originates from
propagator enhancement, favoring high-luminosity facilities.
Away from resonance, however, the leading contribution arises
from interference terms scaling approximately as
$s/M_{Z_\ell}^2$,
which enhances the sensitivity of high-energy colliders.

As a result, circular colliders predominantly probe the
low-mass and small-coupling region,
while linear colliders progressively extend the reach toward
heavier mediator masses.
The transition between these regimes is smooth,
indicating that no single collider configuration can fully
cover the parameter space on its own.

An important implication is that a coherent experimental
program combining circular and linear collider technologies
would maximize the overall sensitivity to leptophilic gauge
interactions.
A signal observed at a high-luminosity machine could,
for example, be further explored at a higher-energy collider,
allowing a more detailed study of the underlying interference
structure and resonance behavior.

Finally, we emphasize that the complementarity observed here
is not tied to the specific benchmark choices adopted in this work,
but instead reflects a generic feature of
interference-sensitive observables at future lepton colliders.

\section{Results and Discussion}
\label{sec:results}

We have presented a phenomenological study of leptophilic gauge
interactions at future $e^+e^-$ colliders, with particular emphasis
on interference-sensitive angular observables and realistic collider effects.

The analysis demonstrates that, over a broad region of parameter space,
the dominant sensitivity arises from interference-induced distortions
of angular distributions rather than from purely resonant rate enhancements.
This behavior is illustrated in Fig.~\ref{fig:angular}
and Fig.~\ref{fig:afb}, where modifications of the differential
distribution generate measurable shifts in the forward--backward asymmetry.

The inclusion of realistic collider effects, such as initial-state
radiation, beamstrahlung, and finite energy resolution,
is essential for obtaining reliable sensitivity projections.
As shown in Fig.~\ref{fig:smearing}, these effects broaden the
resonance structure and reduce the observable peak height relative
to idealized parton-level expectations.

Combining the observable-level analysis with the statistical framework
developed in Sec.~\ref{sec:statistics}, we derived sensitivity projections in the
$(M_{Z_\ell}, g_\ell)$ plane, shown in Fig.~\ref{fig:reach}.
The resulting contours demonstrate that future lepton colliders
can probe leptophilic interactions beyond the parameter region
currently constrained by LEP-inspired bounds.

Several previous studies have investigated leptophilic gauge bosons
using total-rate observables or resonance-based search strategies~\cite{Suarez2025}.
In contrast, the present framework emphasizes the role of
interference-sensitive angular observables in the off-shell regime,
providing a complementary perspective on future collider sensitivity.

The comparison between collider facilities further illustrates the
complementarity between high-luminosity circular colliders and
high-energy linear colliders.
Circular machines are particularly sensitive to small couplings
near the resonance region, while linear colliders extend the accessible
mass reach through the energy growth of interference effects.

Overall, the present study provides an experimentally grounded framework
for probing leptophilic gauge interactions through angular observables
at future lepton colliders.
Future extensions could include detector-level likelihood analyses,
beam-polarization effects, and studies of additional final states such
as $\tau^+\tau^-$ and neutrino channels.

\section{Conclusion}
\label{sec:conclusion}

We have studied the phenomenology of leptophilic gauge
interactions at future $e^+e^-$ colliders within an
observable-driven framework focused on angular observables
and interference effects.

Our analysis shows that, in the off-shell regime,
the dominant sensitivity to a leptophilic gauge boson
is governed primarily by the interference between the
Standard Model and leptophilic amplitudes, leading to
characteristic distortions of the angular distribution.
In this context, the forward--backward asymmetry provides
a robust and experimentally accessible probe of the leading
interference-sensitive component of the signal.

To characterize the relative angular response independently
of the overall rate normalization, we additionally introduced
the interference-normalized quantity $\mathcal{I}_{\rm FB}$.
This quantity provides a useful diagnostic characterization
of the interference structure.

Realistic collider effects, including initial-state radiation,
beamstrahlung, and finite energy resolution, were consistently incorporated.
We showed that these effects can significantly modify resonance profiles
and projected sensitivities, emphasizing the importance of realistic
collider modeling in phenomenological studies of future lepton facilities.

The resulting sensitivity projections in the $(M_{Z_\ell}, g_\ell)$ plane
demonstrate the complementarity between different collider configurations.
High-luminosity circular colliders are particularly sensitive to small
couplings near the resonance region, while high-energy linear colliders
extend the accessible mass reach through the energy growth of
interference effects.

Overall, the present study provides an experimentally grounded framework
for probing leptophilic gauge interactions through interference-sensitive
angular observables at future lepton colliders.
Future extensions based on fully differential angular analyses and
detector-level likelihood methods could further refine the projected sensitivity.

\appendix
\section{Illustrative angular-bin comparison}
\label{app:angular_bins}

The main sensitivity analysis in this work is based on the
forward--backward asymmetry, which compresses the angular distribution
into two hemispheric event yields.
As discussed in Sec.~\ref{sec:statistics}, this choice is motivated by the
fact that the leading off-shell contribution of a heavy leptophilic gauge
boson enters through the interference between the Standard Model
$\gamma+Z$ amplitudes and the $Z_\ell$ exchange amplitude.
This interference term generates a characteristic forward--backward
deformation of the angular distribution.

To illustrate the connection between the integrated asymmetry and the
underlying angular information, we consider a simple binned angular
comparison in $\cos\theta$.
The angular range is divided into four bins,
\[
[-1,-0.5],\qquad [-0.5,0],\qquad [0,0.5],\qquad [0.5,1],
\]
and the normalized bin deviation is defined as
\begin{equation}
\Delta_i^{\rm norm}
=
\frac{
\left(N_i^{\rm SM+Z_\ell}-N_i^{\rm SM}\right)
}{
N_i^{\rm SM}
}.
\label{eq:angular_bin_deviation}
\end{equation}
This quantity is not used as an additional discovery statistic in the
present work.
Rather, it provides a diagnostic check of how the interference-induced
distortion is distributed across the angular bins.

\begin{figure}[t]
\centering

\includegraphics[width=0.82\textwidth]{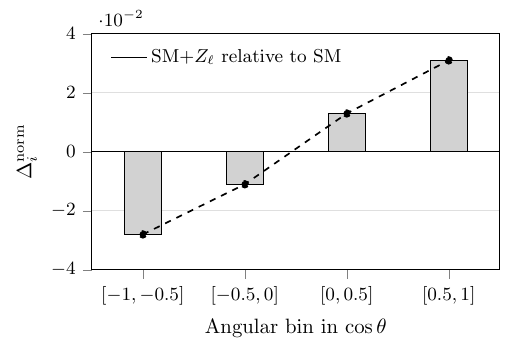}

\caption{
Illustrative normalized angular-bin deviation for a representative
off-shell benchmark point in the interference-dominated regime.
The antisymmetric pattern between the backward and forward hemispheres
shows that the dominant new-physics effect arises from an angular
deformation rather than from an overall normalization shift.
This provides a diagnostic validation of the use of the
forward--backward asymmetry as a compact projection of the leading
interference-sensitive information.
}
\label{fig:angular_bins}

\end{figure}

Figure~\ref{fig:angular_bins} shows an illustrative comparison for a
representative off-shell benchmark point.
The deviation changes sign between the backward and forward hemispheres,
reflecting the antisymmetric angular structure associated with the
interference contribution.
Consequently, the dominant effect is efficiently projected onto
$A_{\rm FB}$, while the binned distribution confirms that the sensitivity
does not arise from a purely normalization-like rate shift.

A fully optimized analysis based on a multi-bin likelihood, including
detector-level correlations and bin-dependent systematic uncertainties,
could further improve the projected reach, but is beyond the scope of the
present phenomenological study.

\FloatBarrier

\bibliographystyle{apsrev4-2}
\bibliography{refs}

\end{document}